\documentclass[preprint,12pt,3p]{elsarticle}

\makeatletter\if@twocolumn\PassOptionsToPackage{switch}{lineno}\else\fi\makeatother

\usepackage{mathtools}

\usepackage{tabulary,xcolor}
\usepackage{amsfonts,amsmath,amssymb}
\usepackage[T1]{fontenc}
\makeatletter
\let\save@ps@pprintTitle\ps@pprintTitle
\def\ps@pprintTitle{\save@ps@pprintTitle\gdef\@oddfoot{\footnotesize\itshape \null\hfill\today}}
\def\hlinewd#1{%
  \noalign{\ifnum0=`}\fi\hrule \@height #1%
  \futurelet\reserved@a\@xhline}

\AtBeginDocument{\ifNAT@numbers \biboptions{sort&compress}\fi}
\makeatother

\usepackage{ifluatex}
\ifluatex
\usepackage{fontspec}
\defaultfontfeatures{Ligatures=TeX}
\usepackage[]{unicode-math}
\unimathsetup{math-style=TeX}
\else 
\usepackage[utf8]{inputenc}
\fi 
\ifluatex\else\usepackage{stmaryrd}\fi

\usepackage{url,multirow,morefloats,floatflt,cancel,tfrupee}
\makeatletter

\AtBeginDocument{\@ifpackageloaded{textcomp}{}{\usepackage{textcomp}}}
\makeatother
\usepackage{colortbl}
\usepackage{xcolor}
\usepackage{pifont}
\usepackage[nointegrals]{wasysym}
\makeatletter

\def\mcWidth#1{\csname TY@F#1\endcsname+\tabcolsep}

\def\cAlignHack{\rightskip\@flushglue\leftskip\@flushglue\parindent\z@\parfillskip\z@skip}
\def\rAlignHack{\rightskip\z@skip\leftskip\@flushglue \parindent\z@\parfillskip\z@skip}

\@ifundefined{etal}{}{}

\usepackage{ifxetex}
\ifxetex\else\if@twocolumn\@ifpackageloaded{stfloats}{}{\usepackage{dblfloatfix}}\fi\fi

\AtBeginDocument{
\expandafter\ifx\csname eqalign\endcsname\relax
\def\eqalign#1{\null\vcenter{\def\\{\cr}\openup\jot\m@th
  \ialign{\strut$\displaystyle{##}$\hfil&$\displaystyle{{}##}$\hfil
      \crcr#1\crcr}}\,}
\fi
}

\AtBeginDocument{%
  \@ifpackageloaded{endfloat}%
   {\renewcommand\efloat@iwrite[1]{\immediate\expandafter\protected@write\csname efloat@post#1\endcsname{}}}{\newif\ifefloat@tables}%
}%

\def\BreakURLText#1{\@tfor\brk@tempa:=#1\do{\brk@tempa\hskip0pt}}
\let\lt=<
\let\gt=>
\def\processVert{\ifmmode|\else\textbar\fi}

\@ifundefined{subparagraph}{
\def\subparagraph{\@startsection{paragraph}{5}{2\parindent}{0ex plus 0.1ex minus 0.1ex}%
{0ex}{\normalfont\small\itshape}}%
}{}

\newcommand\role[1]{\unskip}
\newcommand\aucollab[1]{\unskip}
  
\@ifundefined{tsGraphicsScaleX}{\gdef\tsGraphicsScaleX{1}}{}
\@ifundefined{tsGraphicsScaleY}{\gdef\tsGraphicsScaleY{.9}}{}
\def\checkGraphicsWidth{\ifdim\Gin@nat@width>\linewidth
	\tsGraphicsScaleX\linewidth\else\Gin@nat@width\fi}

\def\checkGraphicsHeight{\ifdim\Gin@nat@height>.9\textheight
	\tsGraphicsScaleY\textheight\else\Gin@nat@height\fi}

\def\fixFloatSize#1{}
\let\ts@includegraphics\includegraphics

\def\inlinegraphic[#1]#2{{\edef\@tempa{#1}\edef\baseline@shift{\ifx\@tempa\@empty0\else#1\fi}\edef\tempZ{\the\numexpr(\numexpr(\baseline@shift*\f@size/100))}\protect\raisebox{\tempZ pt}{\ts@includegraphics{#2}}}}

\AtBeginDocument{\def\includegraphics{\@ifnextchar[{\ts@includegraphics}{\ts@includegraphics[width=\checkGraphicsWidth,height=\checkGraphicsHeight,keepaspectratio]}}}

\DeclareMathAlphabet{\mathpzc}{OT1}{pzc}{m}{it}

\def\URL#1#2{\@ifundefined{href}{#2}{\href{#1}{#2}}}

\def\UrlOrds{\do\*\do\-\do\~\do\'\do\"\do\-}%
\g@addto@macro{\UrlBreaks}{\UrlOrds}

\edef\fntEncoding{\f@encoding}

\makeatother

\newif\ifmultipleabstract\multipleabstractfalse%
\usepackage{amssymb}

\usepackage{subfig}
\usepackage{multirow}
\usepackage{booktabs}

\journal{Nuclear Physics B}

\begin{document}

\begin{frontmatter}
	
\title{\textbf{An encoder-decoder-based method for COVID-19 lung infection segmentation} \textbf{\space }
}

\author[a]{Omar Elharrouss*}
\ead{Elharrouss.omar@gmail.com}

\author[a]{Noor Almaadeed }
\ead{n.alali@qu.edu.qa}

\author[a]{Nandhini Subramanian}
\ead{nandhini.reborn@gmail.com}

\author[a]{Somaya Al-Maadeed}
\ead{s\_alali@qu.edu.qa}

\address[a]{Department of Computer Science and Engineering, Department of Computer Science and Engineering\unskip, 
    Qatar University \unskip, Doha, Qatar\unskip, Doha\unskip, Qatar}

\begin{abstract}

The novelty of the COVID-19 disease and the speed of spread has created a colossal chaos, impulse among researchers worldwide to exploit all the resources and capabilities to understand and analyze characteristics of the coronavirus in term of the ways it spreads and virus incubation time. For that, the existing medical features like CT and X-ray images are used. For example, CT-scan images can be used for the detection of lung infection. But the challenges of these features such as the quality of the image and infection characteristics limitate the effectiveness of these features. Using artificial intelligence (AI) tools and computer vision algorithms, the accuracy of detection can be more accurate and can help to overcome these issues. This paper proposes a multi-task deep-learning-based method for lung infection segmentation using CT-scan images. Our proposed method starts by segmenting the lung regions that can be infected. Then, segmenting the infections in these regions. Also, to perform a multi-class segmentation the proposed model is trained using the two-stream inputs. The multi-task learning used in this paper allows us to overcome shortage of labeled data. Also, the multi-input stream allows the model to do the learning on many features that can improve the results. To evaluate the proposed method, many features have been used. Also, from the experiments, the proposed method can segment lung infections with a high degree performance even with shortage of data and labeled images. In addition, comparing with the state-of-the-art method our method achieves good performance results.

\end{abstract}
\begin{keyword} 
    Lung infection segmentation  \sep  COVID-19  \sep  CT-scan image \sep  Encoder-decoder network 
\end{keyword}

\end{frontmatter}

\section{Introduction}
During the first two months, the coronavirus COVID-19  affected thousands of people around the world with a big number of deaths. Wuhan was the first epicenter of the coronavirus, then it began to spread to every continent and most of the countries \cite{b1}. The severe acute respiratory syndrome or the acronym known by SARS-CoV-2 is an infectious disease that appeared in late 2019. SARS-CoV-2 is known also by COVID-19 due to its similarity to solar corona from the electron microscopic analysis \cite{b2}. The novelty of the Coronavirus and the speed of spread, has created a colossal chaos, impulse among the researchers worlwide to exploit all resources and capabilities to understand and analyze characteristics of the coronavirus in term of spread ways and virus incubation time as well as the exploration of new technics and imposing some temporal procedures to stop the spread speed. The spreading of the virus became unstoppable which prompted the governments to mitigate the impacts of the pandemic using many decisions like stopping the fight and closing the borders and social distancing. However, these policies were not efficient in controlling the spread. To contribute to the implementation of these policies scientists and technology experts attempted to find solutions to stop the speed of the spread. From the technologies used, we can find robots which are used to mitigate the contact between the coronavirus patients and the hospital employees \cite{b3}. Also, drones used for monitoring the social distancing and disinfect the public spaces. 
For the scientists, especially health scientist, searching for medicines and cures that can be helpful to save lives, is the most important mission for them. Whereas the researches in the other domains like computer sciences involved themselves in discovering technics that detect infected persons using the existing medical features like CTscans and X-ray images. Using artificial intelligence (AI) tools and computer vision algorithms, the accuracy of detection can be more accurate and with high precision \cite{bb31,bb32,bb33,bb34,b40,b41}. For that, AI technics can be a good assistant to mitigate the spread by the early detection of the disease. Also, it can be another choice besides the laboratory analysis and allows us to make a large number of tests.

Image processing as well as computer vision combined with AI is a multidisciplinary domain that can be used in different domains including medical, astronomy, agriculture \cite{b4, bb1,bb2, bb3, bb5}. For the medical field, medical imaging has been used for diagnosing diseases using X-ray and CT images, also for surgery and therapy\cite{b3}. Good progress in this field has been reached due to the introduction of many technics like machine learning and deep learning algorithms. This improvement made the computer vision scientists to contribute to finding solutions for rapid diagnostics, prevention, and control. For the same purpose, several approaches have been proposed even when the time is very short.

The detection of infection is the first step for the diagnostic of a disease. Using CT images, it can be seen that the appearances of the infected regions are different from the normal regions, so the detection and the extraction of this region automatically can help the doctor for diagnosis in a short time\cite{b4}. For the same purpose, a deep learning method for segmentation lung infection for COVID-19 on CT images is proposed. Before starting the learning process, each image has been composed of Structure and texture components. The structure represents the homogeneous part of the images whereas the texture component represented the texture of the image. The structure and texture features are introduced to the encoder-decoder model first for segmenting the region of interest, regions that can be affected. Then, segmenting the specific infected parts on the results of the first segmentation. 

 The remainder of the paper is organized as follows. The literature overview is presented in section 2. The proposed method is presented in section 3. Experiments performed to validate the proposed method are discussed in section 4. The conclusion is provided in section 5.   

\begin{table}[t!]
\caption{{Categories of COVID-19 methods} }
\label{tw-150959b0b71c}
\def\arraystretch{1}
\footnotesize
\ignorespaces 
\centering 
\begin{tabulary}{\linewidth}{p{\dimexpr.200\linewidth}p{\dimexpr.27\linewidth-2\tabcolsep}p{\dimexpr.4\linewidth-2\tabcolsep}}
\hline  \textbf{Task} & \textbf{Method} &  \textbf{Architecture} \\
\hline 
\multicolumn{1}{p{\dimexpr(.1602\linewidth-2\tabcolsep)}}{\multirow{11}{\linewidth}{Classification}} &

Ozturk et al.  \cite{bb1} &	DarkNet, CNN\\\cline{2-2}\cline{3-3} &
Minaee et al.  \cite{bb2} &		 ResNet18, CNN\\\cline{2-2}\cline{3-3}&
Apostolopoulos et al.  \cite{bb3} &		 MobileNet, CNN\\\cline{2-2}\cline{3-3}&
Apostolopoulos et al.  \cite{bb4} &		 ResNet18, VGG19, Inception, Xception\\\cline{2-2}\cline{3-3}&
Abbas et al.  \cite{bb5} &		 ResNet18, CNN\\\cline{2-2}\cline{3-3}&

Adhikari et al.  \cite{b5} &		DenseNet-based CNN\\\cline{2-2}\cline{3-3}&

Mobiny et al. \cite{b7} &	CNN\\\cline{2-2}\cline{3-3}&
Polsinelli et al. \cite{b8}  & 	SqueezeNet-based CNN\\\cline{2-2}\cline{3-3}&
Al-karawi et al.\cite{b9} & 	Machine learning, FFT-Gabor \\\cline{2-2}\cline{3-3}&
He et al. \cite{b11}	 &	CNN\\\cline{2-2}\cline{3-3}&

Talha et al.\cite{b12} 	 & 	CNN\\\cline{1-1}\cline{2-2}\cline{3-3}

\multicolumn{1}{p{\dimexpr(.1602\linewidth-2\tabcolsep)}}{\multirow{2}{\linewidth}{Class \& segmentation }} &

Wu et al.  \cite{b6}	 & 	CNN, encoder-decoder
\\\cline{2-2}\cline{3-3}
 &
Amyar et al. \cite{b11} &	CNN, encoder-decoder
\\\cline{1-1}\cline{2-2}\cline{3-3}

\multicolumn{1}{p{\dimexpr(.1602\linewidth-2\tabcolsep)}}{\multirow{1}{\linewidth}{Segmentation  }} &
Fan et al. \cite{b13}	 & CNN, partial decoder
\\\cline{1-1}\cline{2-2}\cline{3-3}
\hline 
\end{tabulary}\par 
\end{table}

\section{Related Works}

Recently, medical imaging has gained attention due to its importance to diagnose, monitor, and treat several medical problems. Radiography, a medical imagining technique, uses \cite{d1,d2,d3}, CT-scan\cite{d4,d5}, and gamma rays to create images of the body that requires internal viewing.  analyzing images using different computer vision algorithms provides an alternative for rapid diagnostics and control of many diseases. For COVID-19, image analysis offers a good solution for early detection due to the complexity of laboratory analysis and the importance of early detection that can save lives. For the same purpose, many approaches have been proposed to COVID-19 detection and control using X-ray and CT-scans images of the lung. Also, the use of artificial intelligence makes the precision and the performance of these results convincing. authors in \cite{b5} presented COVID-Net architecture that is based on DenseNet to diagnose the COVID-19 infections from X-rays and CT-scans images to decrease the turnaround time of the doctors and check more patients at that point of time. Using the proposed architecture, the infected regions are detected and marked. 

Working on similar radiographic images, authors in \cite{b6} developed a classification and segmentation system for a real-time diagnostic of COVID-19. The presented model combines convolutional neural networks and encoder-decoder networks trained on CT-scans images. The proposed approach succeeds to detect the infected regions with an accuracy of 92\%. In the same context, a learning architecture named Detail-Oriented Capsule Networks (DECAPS) has been proposed in\cite{b7}. To increases model stability, authors use an Inverted Dynamic Routing mechanism model implementation. The proposed approach detection the infected region without segmenting the region edges. The same technique is used in\cite{b8} based on SqueezeNet architecture. The obtained results achieve 89\% for performance accuracy wherein \cite{b7} the accuracy reaches 87\%.
To demonstrate the efficiency and the automatic testes of COVID-19 infection, the authors in \cite{b9} proposed a machine learning scheme method for CT-scan images analysis for COVID-19 patients. based on the FFT-Gabor scheme, the proposed method for predicting the state of the patient in real-time. The performance accuracy achieves in average 95.37\% which represents a convincing rate. 

For developing an accurate method for real-time diagnosing of COVID-19 using CT scan images, deep learning models require a large-scale dataset to train these models, which might be difficult to obtain at this moment. Hence, authors in \cite{b10}, build public CT scans dataset people detected positive for COVID-19. Also, a deep-learning-based method has been proposed for classifying the COVID or NON-COVID images. The approach showed an accuracy of 72\%, which means that it is not a very accurate method for COVID-19 testing,
. A multitask deep learning model to identify COVID-19 patient and segmentation of infected regions from chest CT images has been proposed in \cite{b11}. The authors used an encoder-decoder model for segmentation and perceptron for classification. The proposed method has the same technic of the method proposed in \cite{b6} but using different architectures. The obtained performance accuracy of the proposed method reaches 86\%. With an accuracy of 89\%, EfficientNet a neural network architecture is proposed for the detection of COVID-19 patients using CT-scan images \cite{b12}. For segmenting the infected region in a CT-scan image for COVID-19 patient identification, the authors in \cite{b13} proposed a convolutional neural network (CNN) model with a partial decoder. The proposed model performance achieves an accuracy of 73\% for detecting the infected regions in a CT-scan image. The authors also train other architectures including Unet \cite{b14},Unet++ \cite{b15},Attention-Unet \cite{b16},Gated-Unet \cite{b17}, and Dense-Unet \cite{b18} on the same dataset. 

All the proposed method attempts to develop a timely and effective method for testing the coronavirus patients. The proposed methods can be classified into two general categories: Classification methods \cite{b5,b7,b8,b9} and segmentation \cite{b13} of infected region methods. Some of the presented approaches work on two tasks like \cite{b6,b11}. Table 1 presents the proposed method for each category as well as the architectures used in each one of them.

\section{Proposed method}

In this section the proposed approach for lung infection segmentation is provided. The method starts by splitting the texture and structure component of the image, before starting the training of the proposed model for regions of interest extraction. The extraction of these regions is the operation of separation or segmentation of the regions that can contain the infection which are the intern region of the lung. After the extraction of regions of interest, which is performed using the proposed encoder-decoder network proposed, the output of the previous operation is used by the segmentation model which is the sameused for detecting and segmenting the infected part of the lung. A description is provided for each step including stecture-texture decomposition, region of interest extraction and lung infection segmentation.
\begin{figure}[t!]
\centering\includegraphics[width=0.99\linewidth]{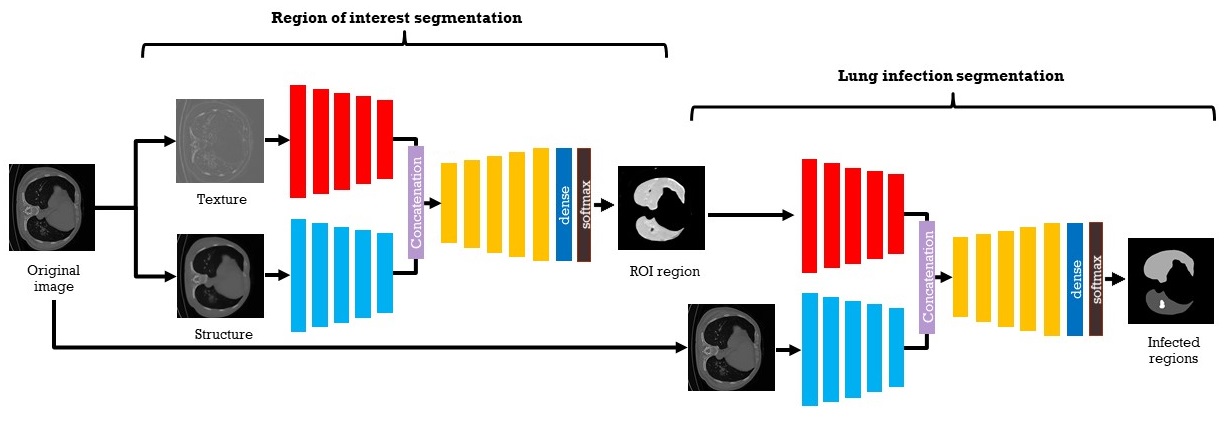}
\caption{Flowchart of the proposed method.}
\end{figure}

\subsection{Structurre and Texture component extraction}

 Each image can contain information that includes the form of the object (or the homogenous part) in the image as well as a texture that also contains information that can be useful in some computer vision tasks. Here, the texture component is used as input of the proposed encoder-decoder neural networks. for that, a preprocessing is performed to extract the texture component. The structure-texture decomposition method proposed in \cite{b30} is adopted using the interval gradient, for adaptive gradient smoothing. Given an image f, it is a technique that splits the image into S+T (f = S+T) of a bounded variation component (Structure) and a component that contains the oscillating part (Texture/Noise) of the image \cite{b291,b292}. We applied this technique on the CT-scan images. Then, we use the texture and structure components as an input of the proposed encoder-decoder model. The extraction of the structure component uses gradient rescaling with interval gradients followed by a color handling operation.

In order to produce a texture-free signal from an input signal I, the gradients within texture regions should be suppressed. Furthermore, the signal should be either increasing or decreasing for all local windows $\Omega_r $ . With these objectives, we use the following equation to rescale the gradients of the input signal with the corresponding interval gradients:

\begin{equation} \label{eq4}
 \begin{array}{@{}l}{\left(\nabla^{'}I\right)}_p=\left\{\begin{array}{lc}{\left(\nabla I\right)}_p.w_p&if\;sign({\left(\nabla I\right)}_p)=sign({\left(\nabla_\Omega I\right)}_p)\\0&otherwise\end{array}\right.\end{array}
\end{equation}

Where$(\nabla^{'}I)_p $ represents the rescaled gradient, and $w_p $ is the rescaling weight:

\begin{equation} \label{eq4}
\begin{array}{@{}l}w_p=min\left(1,\frac{\left|{\left(\nabla_\Omega I\right)}_p\right|+\varepsilon_s}{\left|{\left(\nabla I\right)}_p\right|+\varepsilon_s}\right)\end{array}
\end{equation}

Where $\varepsilon_s $ is a small constant to prevent numerical instability. Too small values of  $\varepsilon_s $ would make the algorithm sensitive to noise, introducing unwanted artifacts to filtering results. The sensitivity to noise can be reduced by increasing \textit{s} but textures may not be completely filtered if  $\varepsilon_s $ is too big. We set  $\varepsilon_s $ = $10^{-4} $ in our implementation.

For filtering color images, we use the gradient sums of color channels in the gradient rescaling step (Equations (1) and (2)), that is:

\begin{equation} \label{eq4}
\begin{array}{@{}l}w_p=min\left(1,\frac{{\displaystyle\sum_{c\in\{r,g,b\}}}\left|{\left(\nabla_\Omega I^{c}\right)}_p\right|+\varepsilon_s}{{\displaystyle\sum_{c\in\{r,g,b\}}}\left|{\left(\nabla I^{c}\right)}_p\right|+\varepsilon_s}\right)\end{array}
\end{equation}

Figure 3 shows filtering examples to demonstrate the results of structure and texture extraction extraction using the same parameters $\varepsilon\in\left[0.01^{2},0.03^{2}\right] $.
\bgroup

\begin{figure}[t!]
  \centering
   \centering\includegraphics[width=0.99\linewidth]{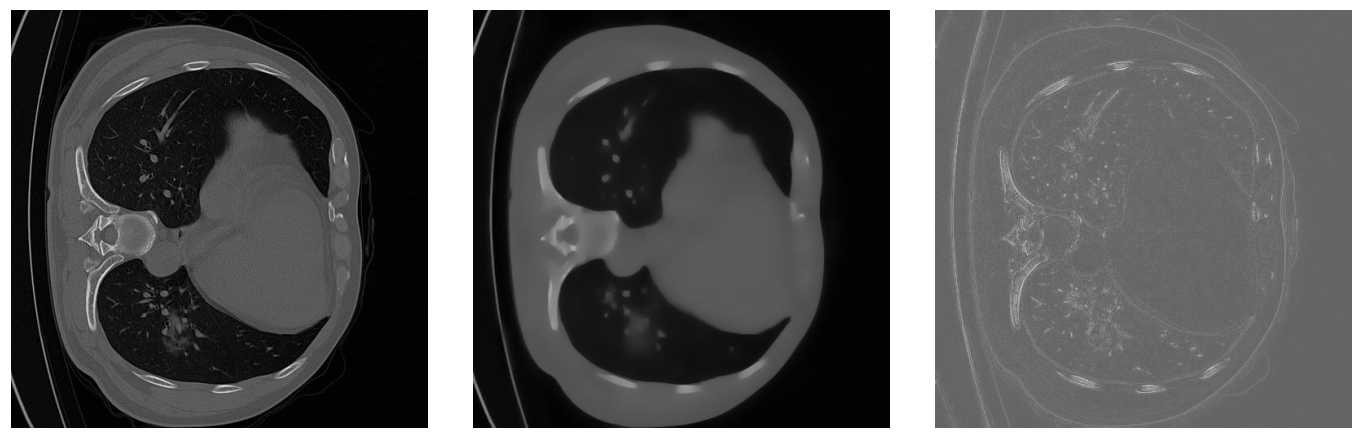}
   \caption{Structure and Texture extraction results: Left: original image. Midle: Structure component. Right: Texture component }
\end{figure}

\subsection{Encoder-decoder architecture}

Dense pixel-wise classification is a required operation for semantic labeling of images. To achieve an effective semantic segmentation, many architectures have been proposed including FCN\cite{b31}, SegNet \cite{b32} architectures. In this paper, we proposed a model that is based on the SegNet model which is an encoder-decoder architecture that provides an image output of the same size as the image input.   

The encoder part of SegNet is based on the VGG-16 \cite{b33} convolutional layers that is composed of 5 blocks, where each one contains 2 to 3 convolutional layers with $3 \times 3$ kernels, 1 padding, ReLU, and a batch normalization (BN)\cite{b34}. The convolution block is followed by a max-pooling layer of size $2 \times 2$. At the end of the encoder, each feature map has $H/32, W/32$, where the original image resolution is $H \times W$.

The decoder part proceeds the operation of upsampling and classification. The main role of this part is the learning of the method of spatial resolution restoration by transforming the encoder features maps into the final labels. The decoder structure is symmetrical with encoder, while the pooling layers in the encoder part are replaced by unpooling layers in the decoder part. The role of convolutional blocks after unpooling layer is to densify the sparse feature maps. This procedure is looped until the feature map reaches the resolution of the input image. 

The final layer in SegNet as well as in most of the proposed architecture in the same context, SoftMax is used to compute the multinomial loss:

\begin{equation} \label{eq4}
   loss = \frac{1}{N}	 \sum_{N=1}^{N} \sum_{i=1}^{k} y_j^{i} log  (\frac{exp(p_j^{i})}{ \sum_{l=1}^{k} exp(p_j^{l})})
\end{equation}

where N is the number of pixels in the input image, k the number of classes and, for a specified pixel $i$; $y_i$ denote its label and PPP the prediction vector. This means that we only minimize the average pixel-wise classification loss without any spatial regularization, as it will be learnt by the network during training. 

In this paper, we followed the same scenarios, but with two streams as input. The encoder part of the proposed model contains two components. The encoder for texture component and the encoder for structure component. the encoder features map is formed by concatenation of these two encoders. In this architecture, we have 5 encoder blocks. Each block of the encoder is composed of a $conv + BN + PReLU + pooling$ where, the decoder blocks composed of $upsampling + conv + BN + PReLU$. The filter size of each convolutional layer in the encoder part is in the range of {32,64, 128, 256}, figure 2 represents the proposed architecture.

\begin{figure}[t!]
  \centering
  \footnotesize
  \begin{tabular}[b]{c}
    \includegraphics[width=.9\linewidth]{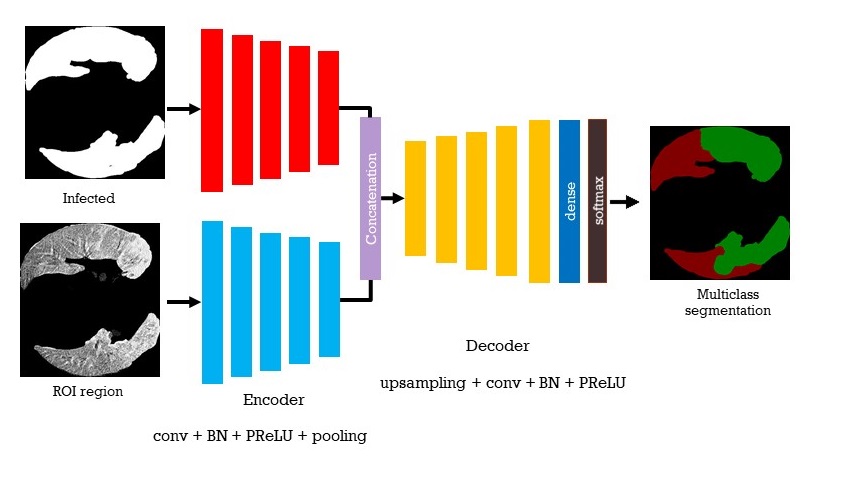} 
  
  \end{tabular}  

   \caption{Multi-class segemrntation of the lung infected regions }
\end{figure}

\begin{figure}[t!]
  \centering
  \footnotesize
\centering\includegraphics[width=0.99\linewidth]{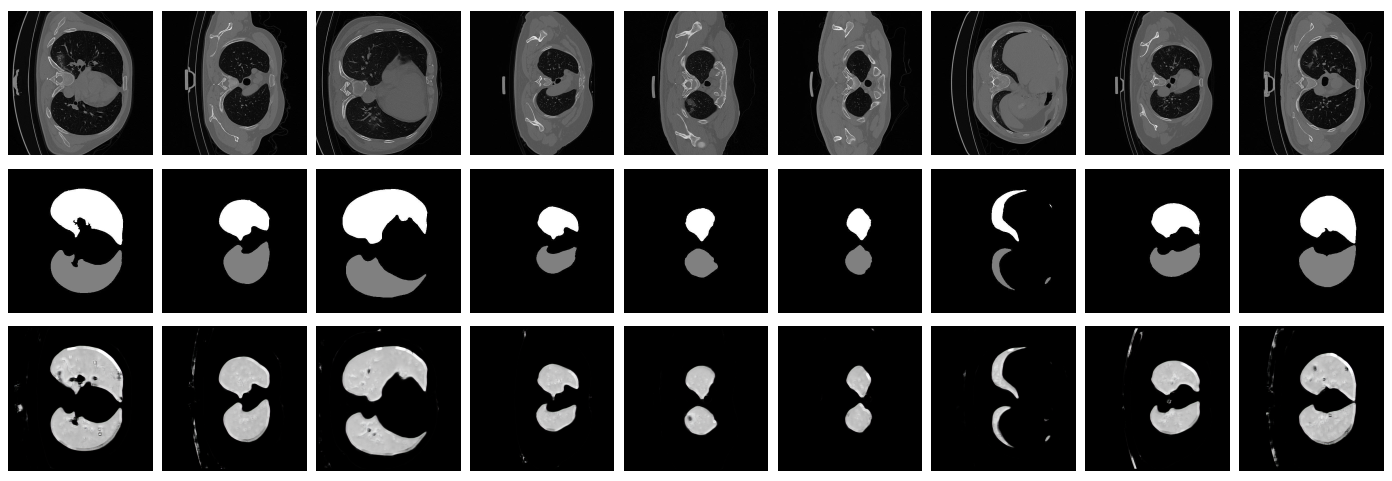}

   \caption{Region of interest segmentation results. First row : original images. Second row: groundtruth. third row: segmentation using the propsoed method.}
\end{figure}
\subsection{Lung infection segmentation }

The current CT-scan dataset of COVID-19 is very limited in terms of the number of labeled images. Also, the manual segmentation of the infected region of the lung is very difficult because that needs a domain expert to do it like a doctor, and also needs time. To solve the current problem of limited data, we augment this data using rotation and translation of a large number of images from the labeled data. using the presented encoder-decoder architecture, we first segment the region of interest or the region that can contain the infection. The inside part of the lung is segmented before used in the next stage of lung infection segmentation. 

For segmenting the lung infected region using black and white colors, where the white color represents the infected region, we use the segmented region which is the output of the first step of segmentation and the original image as input to the model in the second stage as illustrated in figure 1. The multi-task segmentation is an effective solution regarding the limitation of the size of the data discussed above. Also, the segmentation of the infected region can be used for segmenting these regions using multiclass which represents the next task in this paper. 

\subsection{Multi-class lung infection segmentation}

The segmentation using multiclass or many colors to represent the lung infection can be more helpful for the diagnostic of COVID-19 and also more practical. For that, using the previous segmentation of lung infection and the proposed deep learning model, the multiclass segmentation is performed. Two-stream inputs of the deep learning model represented by the lung infection segmentation results and the results of the first segmentation which is the region of interest. This technique allows learning on the specific region and can be more accurate according to the data limitation. Figure 3 illustrates the input and the output of the deep learning model in this stage for multiclass segmentation.

\section{Experimental results}

In this section, we demonstrate the relevance of our proposed method by providing the experimental results. the evaluation has been performed on two segmentation categories including simple segmentation of infected region with black and white, and multi-class labeling using colors. For the first category, we compare our results with a set of state-of-the-art methods, including Unet \cite{b14},Unet++ \cite{b15},Attention-Unet \cite{b16},Gated-Unet \cite{b17}, Dense-Unet \cite{b18} and Semi-Inf-Net \cite{b13}.  For multi-class labeling, the obtained results are compared with four state-of-the-art methods such as Semi-Inf-Net \cite{b13}, multi-class U-Net \cite{b19}, FCN8s \cite{b20} and DeepLabV3+ \cite{b21}. Also, the results are visualized by presenting some segmented examples using the proposed method as well as state-of-the-art methods.

\begin{figure}[t!]
 \centering
\scriptsize
  \begin{tabular}[b]{c}
    \includegraphics[width=.16\linewidth]{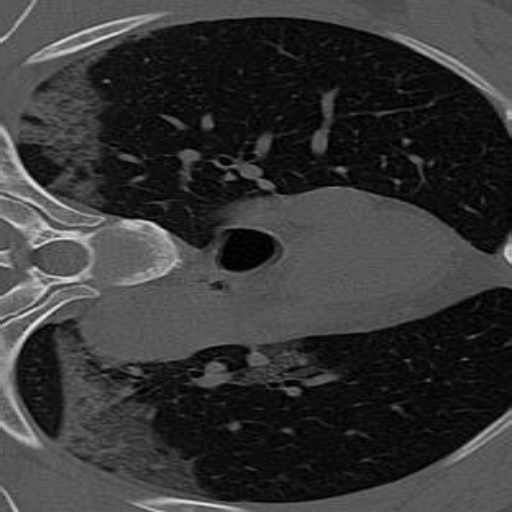} \\
  \end{tabular}
  \begin{tabular}[b]{c}
   \includegraphics[width=.16\linewidth]{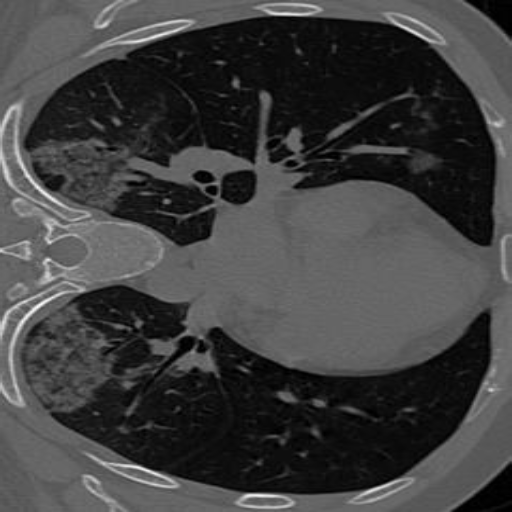}\\
  \end{tabular}
  \begin{tabular}[b]{c}
  \includegraphics[width=.16\linewidth]{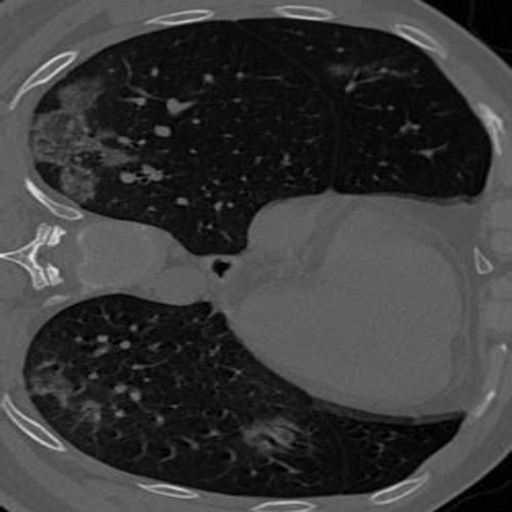}\\
  \end{tabular}
  \begin{tabular}[b]{c}
  \includegraphics[width=.16\linewidth]{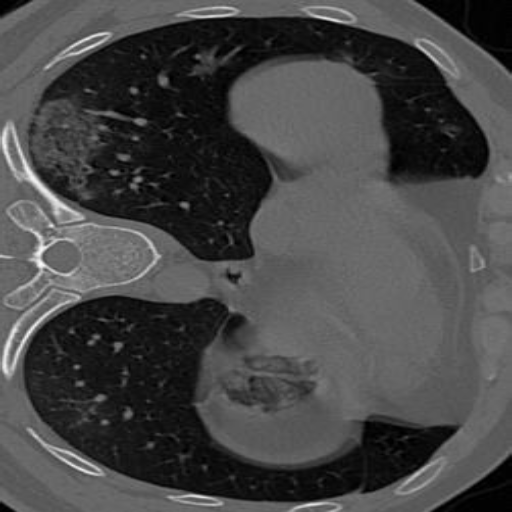}\\
  \end{tabular}
  \begin{tabular}[b]{c}
  \includegraphics[width=.16\linewidth]{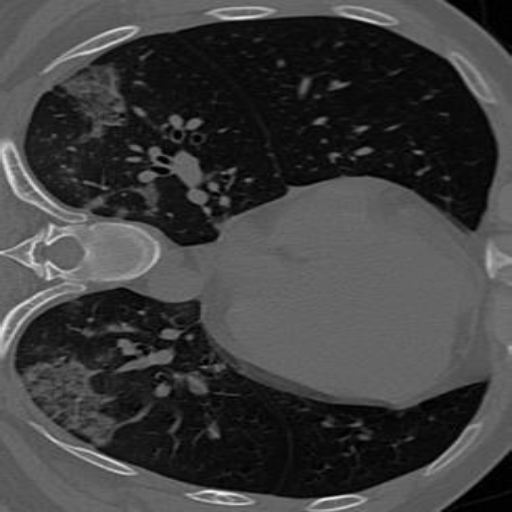}\\
  \end{tabular}
  \begin{tabular}[b]{c}
    \includegraphics[width=.16\linewidth]{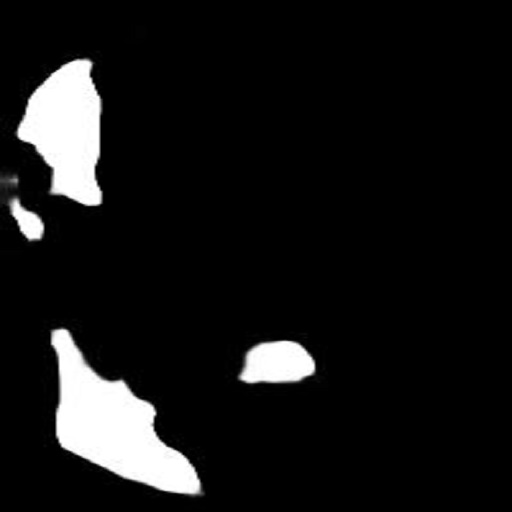} \\
   \end{tabular}
  \begin{tabular}[b]{c}
   \includegraphics[width=.16\linewidth]{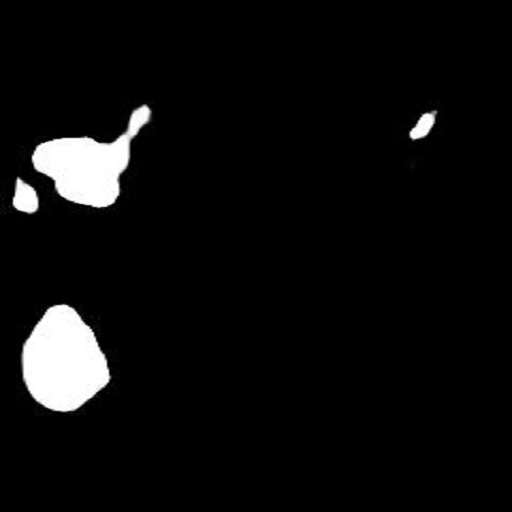}\\
  \end{tabular}
  \begin{tabular}[b]{c}
  \includegraphics[width=.16\linewidth]{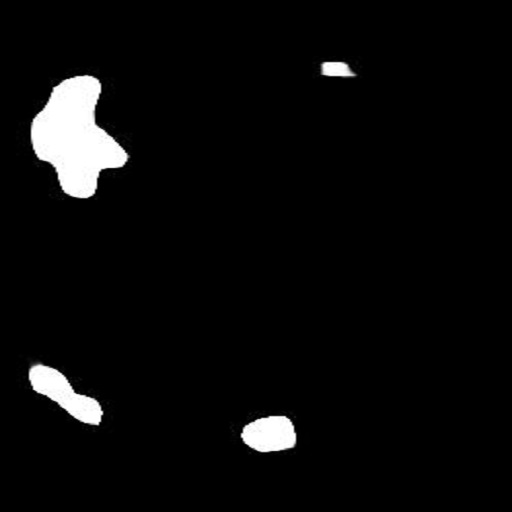}\\
  \end{tabular}
  \begin{tabular}[b]{c}
  \includegraphics[width=.16\linewidth]{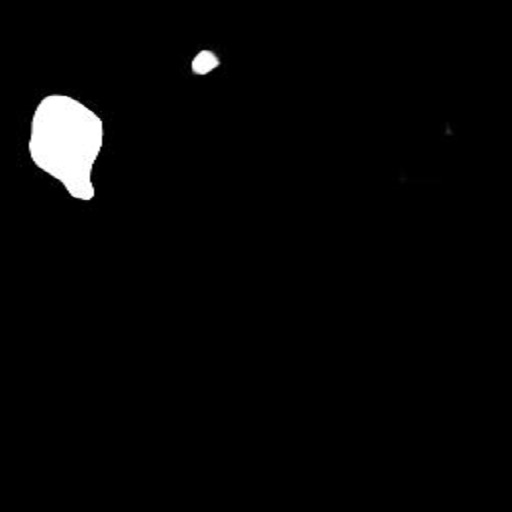}\\
  \end{tabular}
  \begin{tabular}[b]{c}
  \includegraphics[width=.16\linewidth]{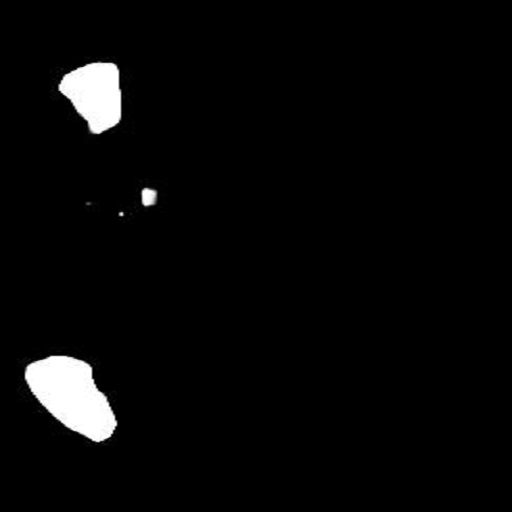}\\
 \end{tabular}
  \begin{tabular}[b]{c}
    \includegraphics[width=.16\linewidth]{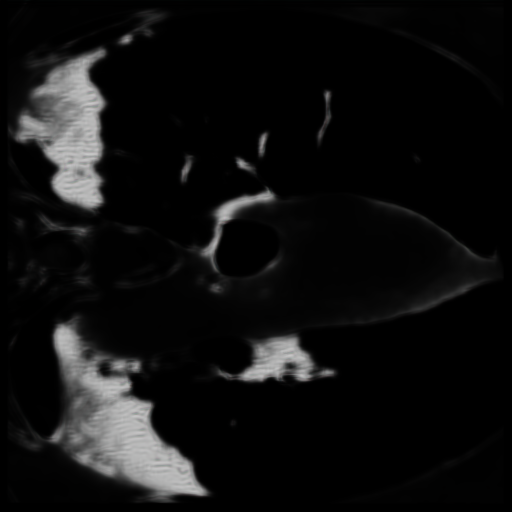} \\
Dice= 0.849\\
Sensitivity=0.834\\
Specitivity=0.987\\
Precision=0.864\\
Fmeasure=0.849\\

 \end{tabular}
  \begin{tabular}[b]{c}
   \includegraphics[width=.16\linewidth]{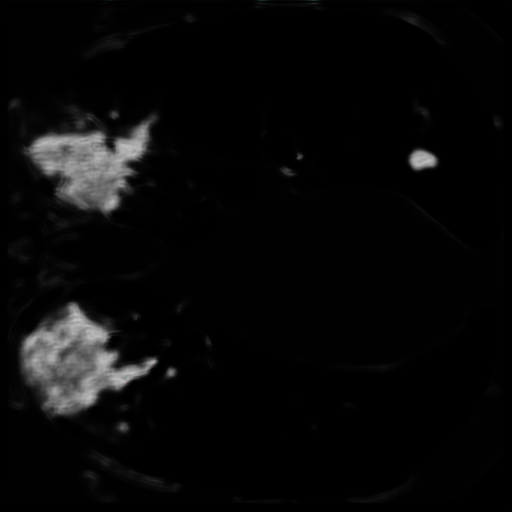}\\
Dice= 0.870\\
Sensitivity=0.749\\
Specitivity=0.990\\
Precision=0.814\\
Fmeasure=0.780\\

 \end{tabular}
  \begin{tabular}[b]{c}
  \includegraphics[width=.16\linewidth]{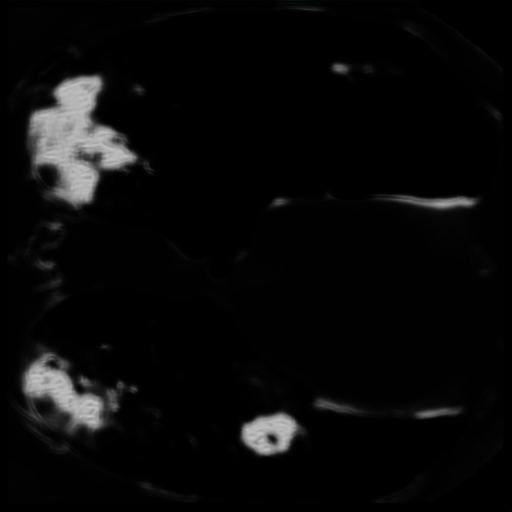}\\
Dice= 0.850\\
Sensitivity=0.794\\
Specitivity=0.996\\
Precision=0.915\\
Fmeasure=0.850\\

 \end{tabular}
  \begin{tabular}[b]{c}
  \includegraphics[width=.16\linewidth]{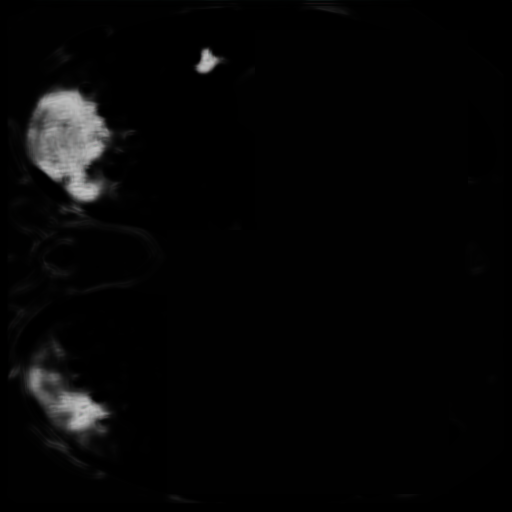}\\
Dice= 0.765\\
Sensitivity=0.651\\
Specitivity=0.998\\
Precision=0.929\\
Fmeasure=0.767\\
 \end{tabular}
  \begin{tabular}[b]{c}
  \includegraphics[width=.16\linewidth]{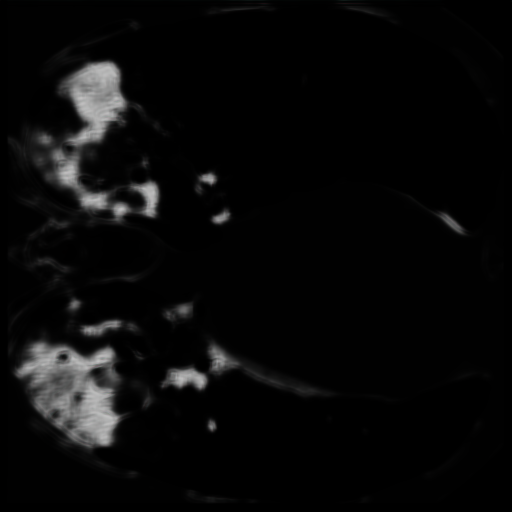}\\
Dice= 0.676\\
Sensitivity=0.529\\
Specitivity=0.997\\
Precision=0.936\\
Fmeasure=0.676\\

 \end{tabular}
 
   \caption{infection region segmentation . First row : original images. Second row: groundtruth. Third row: segmentation using the propsoed method. fourth row: evaluation results }
\end{figure}
\subsection{Datasets}

 The only segmentation dataset of CT-scan images available for COVID-19 is \footnote{https://medicalsegmentation.com/covid19/}. The dataset consists of 100 axial CT images for 20 COVID-19 patients, collected by the Italian Society of Medical and Interventional Radiology. The dataset contains CT images labeled. There are two categories of labels. The first one labels the regions of interest where the infection can be located. The other one is the specific infected regions labeled with two colored red and green colors. This part contains cropped images for training and testing. The training images composed of 50 images where the infections are labeled with one color (one class) and multi-class (2 colors). The test folder contains 48 images labeled with the same labels as the training. in this paper, we train our multitask model on the two categories of images. We train the first part of our model on the 2000 images for region segmentation. For infection segmentation, we use the labeled data and we augment the data by rescaling and rotating the images for getting more data. 

\begin{figure}[t!]
  \centering
  \footnotesize
\centering\includegraphics[width=0.99\linewidth]{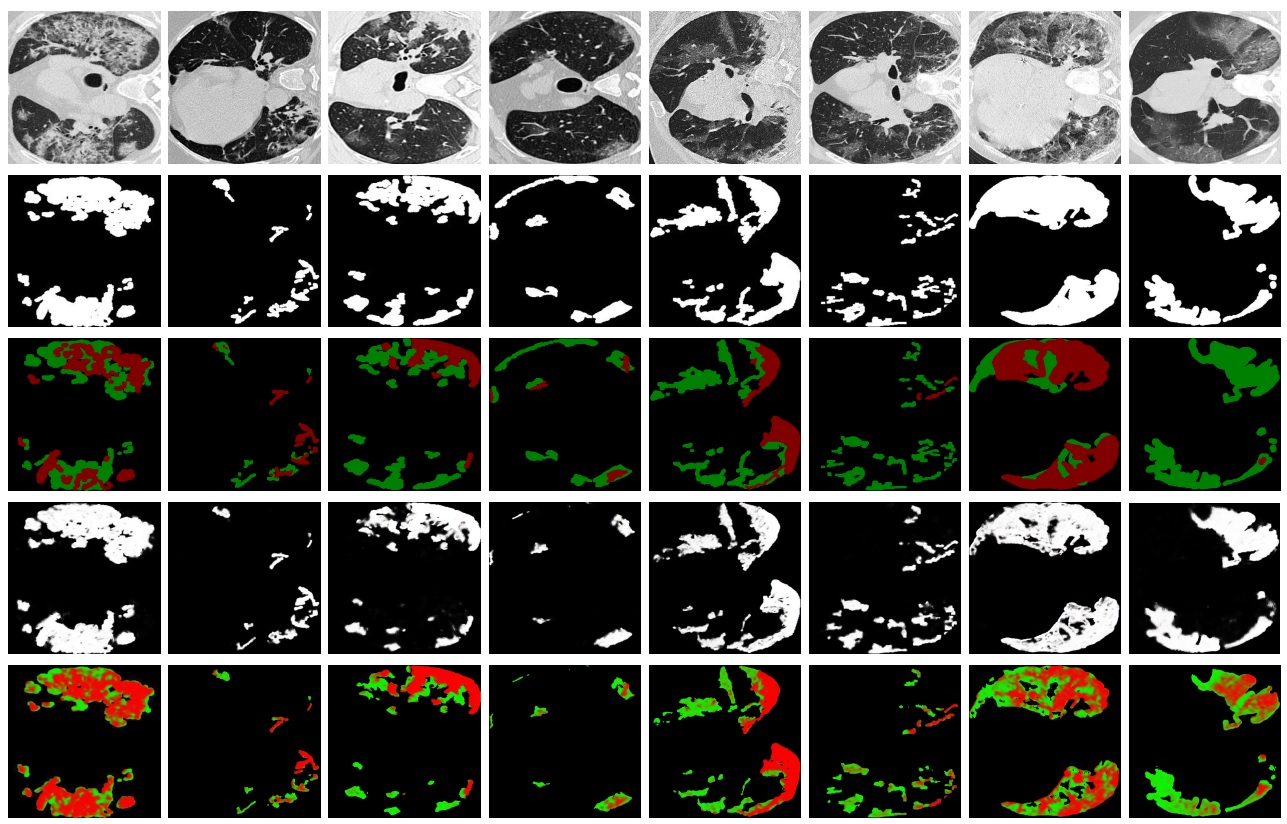}

   \caption{Segmentation results using the proposed method. First row: original images. Second row: binary ground truth. Third row: multi-class ground truth. Fifth row: Our binary segmentation results. Sixth row:  Our multi-class segmentation results.  }
\end{figure}

\subsection{Evaluation metrics}
 
To evaluate the segmentation results of the propsoed method, a set of measures has been exploited including Sorensen-Dice similarity, Sensitivity, Specitivity, Precision and F-measure used in \cite{b361,b37, b38, b39}.  For calculation of these
three parameters, the four measures are required, namely true positive (TP), false positive (FP), true negative (TN), and false negative (FN). The true positive (TP) represents the number of pixels being correctly identifed. True
negative (TN) describe the number of non-lung infection pixels being correctly identifed as non-lung infection. False positive (FP) denotes the number of non-lung infection being wrongly classifed as lung infection, whereas false negative (FN) means the lung infection pixels  being wrongly
classifed as non-lung infecttion.

\textbf{Sorensen-Dice similarity:}
Let consider that $A$ is segmented regions that we need to assess the quality, $B$ is ground truth. The Sorensen-Dice similarity \cite{b37} is computed as follows:
\begin{equation} \label{eq4}
  Dice= \frac{2|A\cap B|}{|A|+ |B|}
\end{equation}

The value of the Sorensen-Dice similarity metric is between 0 and 1. The higher the Sorensen-Dice value, the better the segmentation result.

\textbf{Sensitivity:}
Sensitivity, defned as the ratio of correctly identifed lung infection to the total number of lung infection pixels, is computed as

\begin{equation} \label{eq4}
 Sensitivity= \frac{TP}{TP+ FN}
\end{equation}

\textbf{Specifcity:}
Specifcity, defned as the ratio of correctly detected non-lung unfection to the total number of non-lung infection pixels, is measured as

\begin{equation} \label{eq4}
 Specifcity= \frac{TN}{TN+ FP}
\end{equation}

\textbf{Precision:}
Precision gives the percentage of unnecessary positives through the compared total number of positive pixels in the detected binary objects mask \cite{b39}
\begin{equation} \label{eq4}
 Precision= \frac{TP}{TP+ FP}
\end{equation}

\textbf{ F-measure:}
the
weighted harmonic mean of precision and sensitivity, computes the quality of detection.

\begin{equation} \label{eq4}
  F-measure= \frac{2 \times Sensitivity \times Precision }{Sensitivity + Precision }
\end{equation}

Mean Absolute Error (MAE) is also used to evaluate the performance of the proposed model:

\begin{equation} \label{eq4}
  MAE = \frac{1}{N} \sum_{i=1}^{N} | z_i - z_{i}^{gt}|
\end{equation}

\begin{table}[t!]
\caption{{Quantitative results of binary infection regions segmentation on COVID-SemiSeg DATASET} }
\label{tw-150959b0b71c}
\def\arraystretch{1}
\footnotesize
\ignorespaces 
\centering 
\begin{tabulary}{\linewidth}{p{\dimexpr.250\linewidth}p{\dimexpr.12\linewidth-2\tabcolsep}p{\dimexpr.12\linewidth-2\tabcolsep}p{\dimexpr.12\linewidth-2\tabcolsep}p{\dimexpr.12\linewidth-2\tabcolsep}p{\dimexpr.12\linewidth-2\tabcolsep}p{\dimexpr.12\linewidth-2\tabcolsep}}
\hline Method & Dice & Sensitivity & Specitivity & Precision & Fmeasure & MAE\\
\hline 
U-Net \cite{b14} & 0.439 & 0.534 & 0.858 & -  &   - & 0.186
\\\cline{1-1}\cline{2-2}\cline{3-3}\cline{4-4}\cline{5-5}\cline{6-6}\cline{7-7}
Attention-UNet \cite{b16} &  0.583  & 0.637 & 0.921 & -  &   - & 0.112
\\\cline{1-1}\cline{2-2}\cline{3-3}\cline{4-4}\cline{5-5}\cline{6-6}\cline{7-7}
Gated-UNet \cite{b17} &  0.623 & 0.658 & 0.926& -  &   -  & 0.102
\\\cline{1-1}\cline{2-2}\cline{3-3}\cline{4-4}\cline{5-5}\cline{6-6}\cline{7-7}
Dense-UNet \cite{b18} & 0.515 & 0.594 & 0.840 & -  &   -  & 0.184
\\\cline{1-1}\cline{2-2}\cline{3-3}\cline{4-4}\cline{5-5}\cline{6-6}\cline{7-7}
U-Net++  \cite{b15} & 0.422 & 0.379 & 0.976& -  &   -  & 0.120
\\\cline{1-1}\cline{2-2}\cline{3-3}\cline{4-4}\cline{5-5}\cline{6-6}\cline{7-7}
Semi-Inf-Net \cite{b13} &\textcolor{blue}{ 0.739} &\textcolor{red}{ 0.725} & \textcolor{blue}{0.960} & - &  - & \textcolor{red}{0.064}
\\\cline{1-1}\cline{2-2}\cline{3-3}\cline{4-4}\cline{5-5}\cline{6-6}\cline{7-7}

Propsed method  &	\textcolor{red}{0.786} &	\textcolor{blue}{0.711} & \textcolor{red}{0.993} & \textcolor{red}{0.856} & \textcolor{red}{0.784} & \textcolor{blue}{0.076}
\\\cline{1-1}\cline{2-2}\cline{3-3}\cline{4-4}\cline{5-5}\cline{6-6}\cline{7-7}
\end{tabulary}
\end{table}

\subsection{Discussion}

The first step of our proposed model is to segment the regions that can contain the infection from lung images. This step is sued for segmenting the lung infection regions in the second part of the proposed method. This step is an accurate preprocessing operation for segmenting the lung infection region. Figure 4 illustrates some obtained results for the region of interest segmentation by presenting the original image with the ground truth as well as the obtained results. We can observe that the proposed model gives promising segmentation results. Also, comparing with the ground truth, we obtain a segmentation without fold segmented regions.

To evaluate the performance of the proposed method for lung infection segmentation Figure 5 show some examples of the obtained results. From the visualized results we can observe that the proposed model can detect the infection effectively with some errors that can be considered negligible. Also, the segmentation results are close to the ground truth show in the second line of figure 5. The success of the proposed method to label the infection is owed to the used architecture that uses two-stream input which allow robust learning. Also, the use of the results of the region of interest as input beside the original image gives the model a specification of the region that can contain the infection. The robustness of the presented approach is shown also in the presented metrics that demonstrate the performance evaluation including Dice, Sensitivity, specificity, precision, and F-measure. For example, for the first CT-scan image, our approach can segment with a high precision value which achieves 84\% for the F-measure metric which is a convincing value. The presented results can be improved using a preprocessing on the results images like the morphology operations.

In order to assert the results represented by the binary image in Fig. 5, the qualitative and quantitative results obtained by each method are represented in Table 2. From these results, we can see that the proposed method gives high performance when compared with the state-of-art methods. The effectiveness of the proposed method comes from the use of multi-task learning and the use of two-stream input for our model. 

The multi-class infection labeling results has also been illustrated in Figure 6. As shown in the figure the proposed method performs an accurate segmentation of the lung infection using multi-class labeling. The best result comes from the succession of tasks for performing the multi-class segmentation. The use of the results of unit-class segmentation with the original image leads to a precise segmentation of the lung infection. The evaluation using different metric in  Table 3 also demonstrate the advantage of the proposed method compared with the other existing methods. For example, the semi-inf-Net method succeed to obtain close results due to the multi-task learning model. In contrast to the other model that are used as it is like UNet of FC8s models. 

\begin{table}[t!]
\caption{{Quantitative results of infection regions using multi-class segmentation on COVID-SemiSeg DATASET} }
\label{tw-150959b0b71c}
\def\arraystretch{1}
\footnotesize
\ignorespaces 
\centering 
\begin{tabulary}{\linewidth}{p{\dimexpr.250\linewidth}p{\dimexpr.12\linewidth-2\tabcolsep}p{\dimexpr.12\linewidth-2\tabcolsep}p{\dimexpr.12\linewidth-2\tabcolsep}p{\dimexpr.12\linewidth-2\tabcolsep}p{\dimexpr.12\linewidth-2\tabcolsep}p{\dimexpr.12\linewidth-2\tabcolsep}}
\hline Method & Dice & Sensitivity & Specitivity & Precision & Fmeasure & MAE \\
\hline 
multi-class U-Net \cite{b19} & 0.422 & .0379 & 0.976 & -  &   - & 0.066
\\\cline{1-1}\cline{2-2}\cline{3-3}\cline{4-4}\cline{5-5}\cline{6-6} \cline{7-7}

DeepLabV3+ \cite{b21} & 0.341 & 0.512 & 0.766 & -  &   - & 0.117
\\\cline{1-1}\cline{2-2}\cline{3-3}\cline{4-4}\cline{5-5}\cline{6-6}\cline{7-7}
FC8s \cite{b20} & 0.375 & 0.403 & 0.811  & -  &   -  & 0.076
\\\cline{1-1}\cline{2-2}\cline{3-3}\cline{4-4}\cline{5-5}\cline{6-6}\cline{7-7}

Semi-Inf-Net \cite{b13} & \textcolor{blue}{0.541} & \textcolor{blue}{0.564} & \textcolor{red}{0.967}& - &  - & \textcolor{red}{0.057}
\\\cline{1-1}\cline{2-2}\cline{3-3}\cline{4-4}\cline{5-5}\cline{6-6}\cline{7-7}

Propsed method  &	\textcolor{red}{0.640}  &	\textcolor{red}{0.630}	& \textcolor{blue}{ 0.953 } & \textcolor{red}{0.561} & \textcolor{red}{0.640} & \textcolor{blue}{0.062}
\\\cline{1-1}\cline{2-2}\cline{3-3}\cline{4-4}\cline{5-5}\cline{6-6}\cline{7-7}
\end{tabulary}
\end{table}

\section{Conclusion}

In this paper, a lung infection segmentation method for COVID-19 has been proposed. Based on encoder-decoder networks on CT-scan images exploit the computer vision techniques to identify the lung infected regions for COVID-19 patients. Due to the shortage of data at this moment, multi-task learning has been performed including the use of two-stream as inputs of the deep learning model. Also, using computer vision features like structure and texture component of the images that help for good extraction of the region of interest that can contain infections has also been utilized. Different segmentation has been performed including the binary segmentation and multi-class segmentation of lung infected regions. comparing the proposed approach with the state-of-the-art method, the experiment shows an accurate segmentation of the lung infection region in both binary and multiclass segmentation.  The obtained results can be improved by using more data for training and more labeled data for multi-class segmentation, which represents our future works.


\section{Acknowledgements}

This publication was made by rEdicting RIsk earLy in COVID-19 project  (QUERG-CENG-2020-1). The statements made herein are solely
the responsibility of the authors.

\section{Conflict interests}
The authors declare that they have no competing interests.

\end{document}